\documentclass{icrc2009}

\usepackage{graphicx}   
\usepackage{caption}    
\usepackage[font=footnotesize]{subfig} 
\usepackage{fixltx2e}
\usepackage{url}

\newcommand{\shorttitle}[1]%
{\markboth{Proceedings of the 31\MakeLowercase{$^{st}$} ICRC, {\L}\'{o}d\'{z} 2009}{#1} }
\newcommand{\etal}{\MakeLowercase{\textit{et al. }}} 

\begin{document}
\title{Small air showers in IceTop}

\author{\IEEEauthorblockN{Bakhtiyar Ruzybayev\IEEEauthorrefmark{1},
                          Shahid Hussain\IEEEauthorrefmark{1},
                          Chen Xu\IEEEauthorrefmark{1} and
                          Thomas Gaisser\IEEEauthorrefmark{1} 
                          for the IceCube Collaboration\IEEEauthorrefmark{2}
                          }
                            \\
\IEEEauthorblockA{\IEEEauthorrefmark{1}Bartol Research Institute, Department of Physics and Astronomy, University of Delaware, Newark, DE 19716, U.S.A.}
\IEEEauthorblockA{\IEEEauthorrefmark{2}See the special section of these proceedings.}
}

\shorttitle{B.{~}Ruzybayev\etal Small air showers}
\maketitle

\begin{abstract}
 IceTop is an air shower array that is part of the IceCube Observatory currently under construction at the geographic South Pole \cite{Tom}.
 When completed, it will consist of 80 stations covering an area of 1\,km$^{2}$.
 Previous analyzes done with IceTop studied the events that triggered five or more stations,
 leading to an effective energy threshold of about 0.5\,PeV \cite{fabian}. 
 The goal of this study is to push this threshold lower, into the region where it will overlap with direct measurements of cosmic rays which
 currently have an upper limit around 300\,TeV \cite{pdg}.
 We select showers that trigger exactly three or exactly four adjacent surface stations that are not on the periphery of the detector (contained events).
 This extends the energy threshold down to 150\,TeV.
\end{abstract}


\section{Introduction}
During 2008, IceCube ran with forty IceTop stations and forty IceCube strings in a triangular grid with a mean separation of 125\,m.
In the 2008--2009 season, additional 38 IceTop tanks and 18 standard IceCube strings were deployed as shown in Fig.1. 
When completed, IceCube will consist of eighty surface stations, eighty standard strings and six special strings in the ''DeepCore'' sub-array \cite{icecube}.
Each IceTop station consists of two ice filled tanks separated by 10\,m, each equipped with two Digital Optical Modules (DOMs) \cite{doms}.
The photo multipliers inside the two DOMs are operated at different gains to increase the dynamic range of the response of a tank. The DOMs detect the Cherenkov light 
emitted by charged shower particles inside the ice tanks. Data recording starts when local coincidence condition is satisfied, that is when both tanks are hit within a
250 nanoseconds interval. In this paper we used the experimental data taken with the forty station array and compared to simulations of this detector configuration. Here we
describe the response of IceTop in its threshold region.

\begin{figure}[!t]
  \centering
  \includegraphics[width=3.2in]{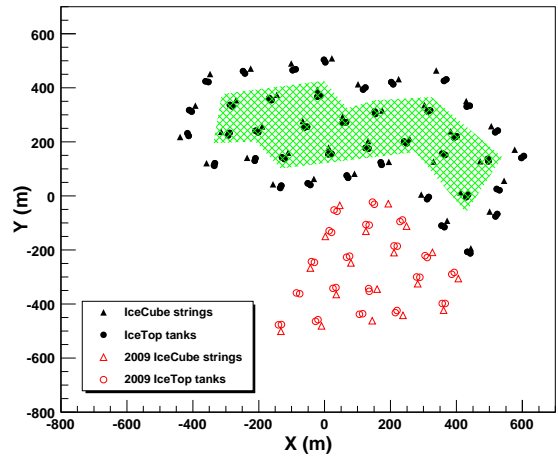}
  \caption{Surface map of IceCube in 2009. New stations are unfilled markers. The shaded area (200795\,m$^{2}$) contains stations that are defined as inner stations.}
  \label{array}
\end{figure}

\section{Analysis}
The main difference between this study and analyzes done with five or more stations 
triggering is the acceptance criterion. In previous analyzes, we accepted events with five or more hit stations and with reconstructed shower core location within the 
predefined containment area (shaded area in Fig.1). In addition, the station with the biggest signal in the event must also 
be located within the containment area.

In the present analysis we used events that triggered only three or four stations, thus complementing analyzes with five or more stations. 
Selection of the events was based solely on the stations that were triggered. The criteria are:

 \begin{enumerate}
 \setlength{\itemsep}{-2pt}
 \item
       Triggered stations must be close to each other (neighboring stations). For three station events, stations form almost an equilateral triangle. 
       For four station events, stations form a diamond shape.  \\
 \item
       Triggered stations must be located inside the array (shaded region in Fig.1). Events that trigger stations on the periphery are discarded.  
 \end{enumerate} 

\begin{figure*}[th]
  \centering
  \includegraphics[width=4.0in]{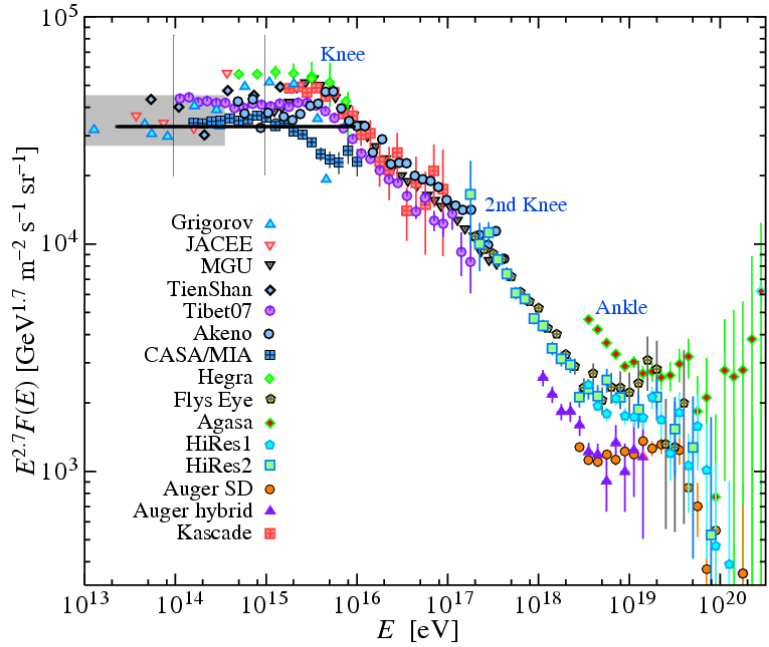}
  \caption{The all-particle spectrum from air shower measurements as summarized in Figure 24.9 of Review of Particle Physics \cite{pdg}.
 The shaded area indicates the range of direct measurements. The thick black line shows the flux model used for this analysis and 
 the vertical lines indicate the energy range responsible for 95\% of the 3 and 4 station events.}
  \label{flux}
 \end{figure*}

Since we are using stations on the periphery as a veto, we ensure that our selected events will have
shower cores contained within the boundary of the array. In addition, these events will have a narrow energy distribution. 
We analyzed events in four solid angle bins with zenith angles $\theta$: 0$^{\circ}$--26$^{\circ}$, 26$^{\circ}$--37$^{\circ}$,
37$^{\circ}$--45$^{\circ}$, 45$^{\circ}$--53$^{\circ}$. Results for the first bin, $\theta = 0^{\circ}$--$26^{\circ}$, are emphasized in this paper.
This near-vertical sample will include most of the events with muons seen in coincidence with the deep part of IceCube.

\begin{figure*}[th]
  \centering
  \includegraphics[width=4.0in]{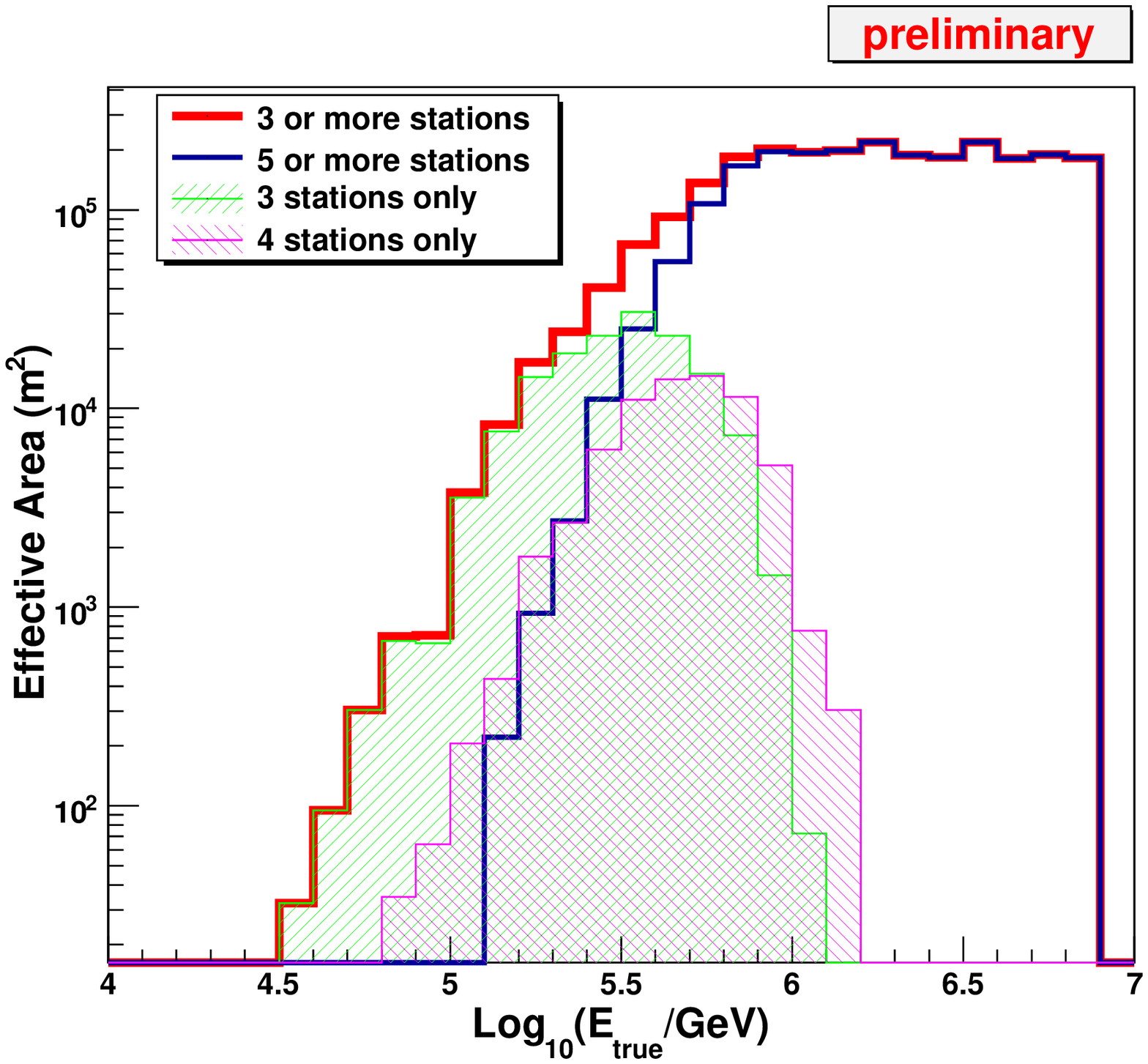}
  \caption{Effective areas for different triggers in the most vertical zenith angle range: $0^{\circ}\leq\theta\leq26^{\circ}$,
           derived using true quantities from simulations.}
  \label{effarea}
 \end{figure*}

\begin{figure*}[!t]
   \centerline{\subfloat[Proton simulation]{\includegraphics[width=3.1in]{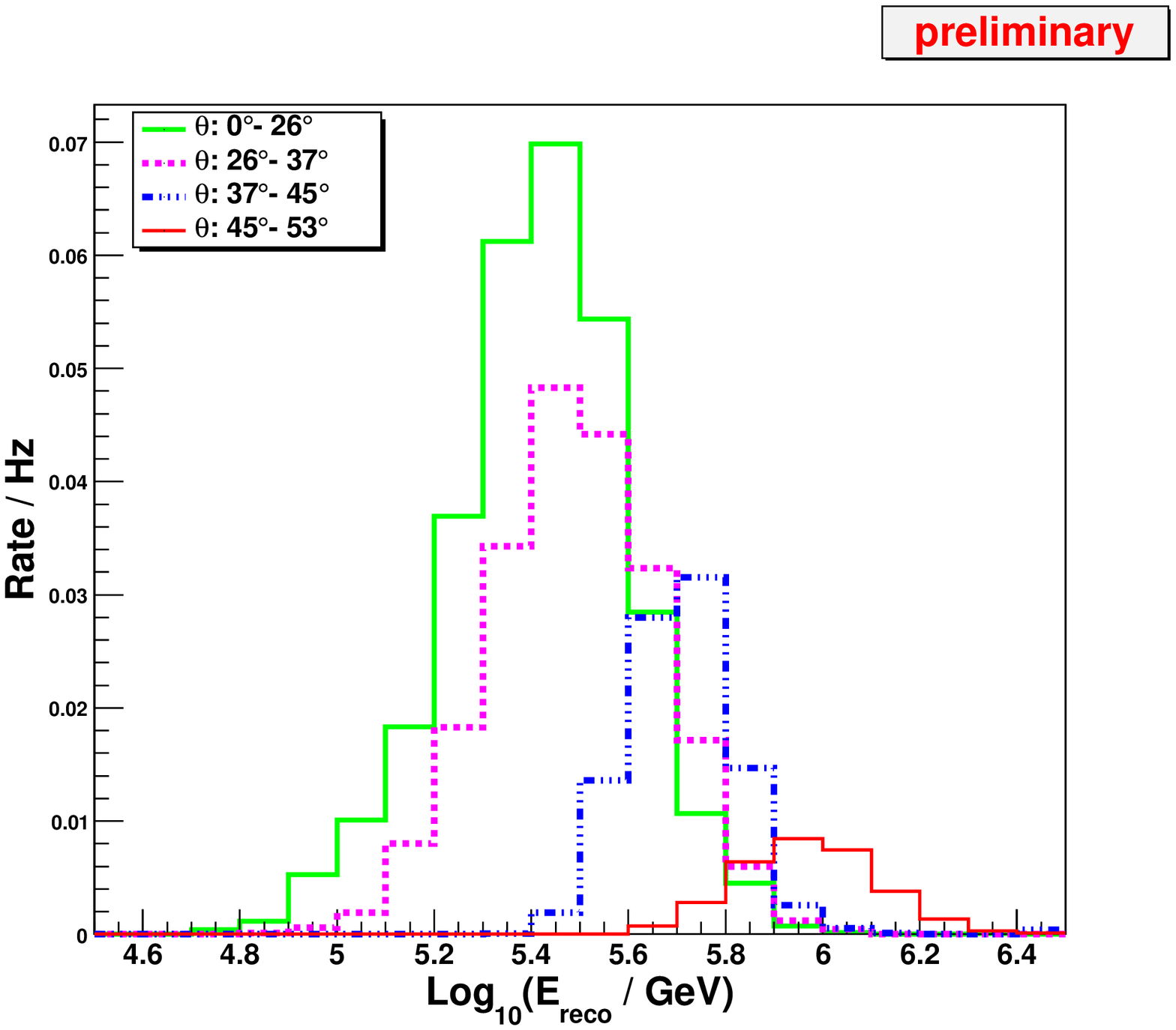} \label{sub_fig1}}
              \hfil
              \subfloat[Iron simulation]{\includegraphics[width=3.1in]{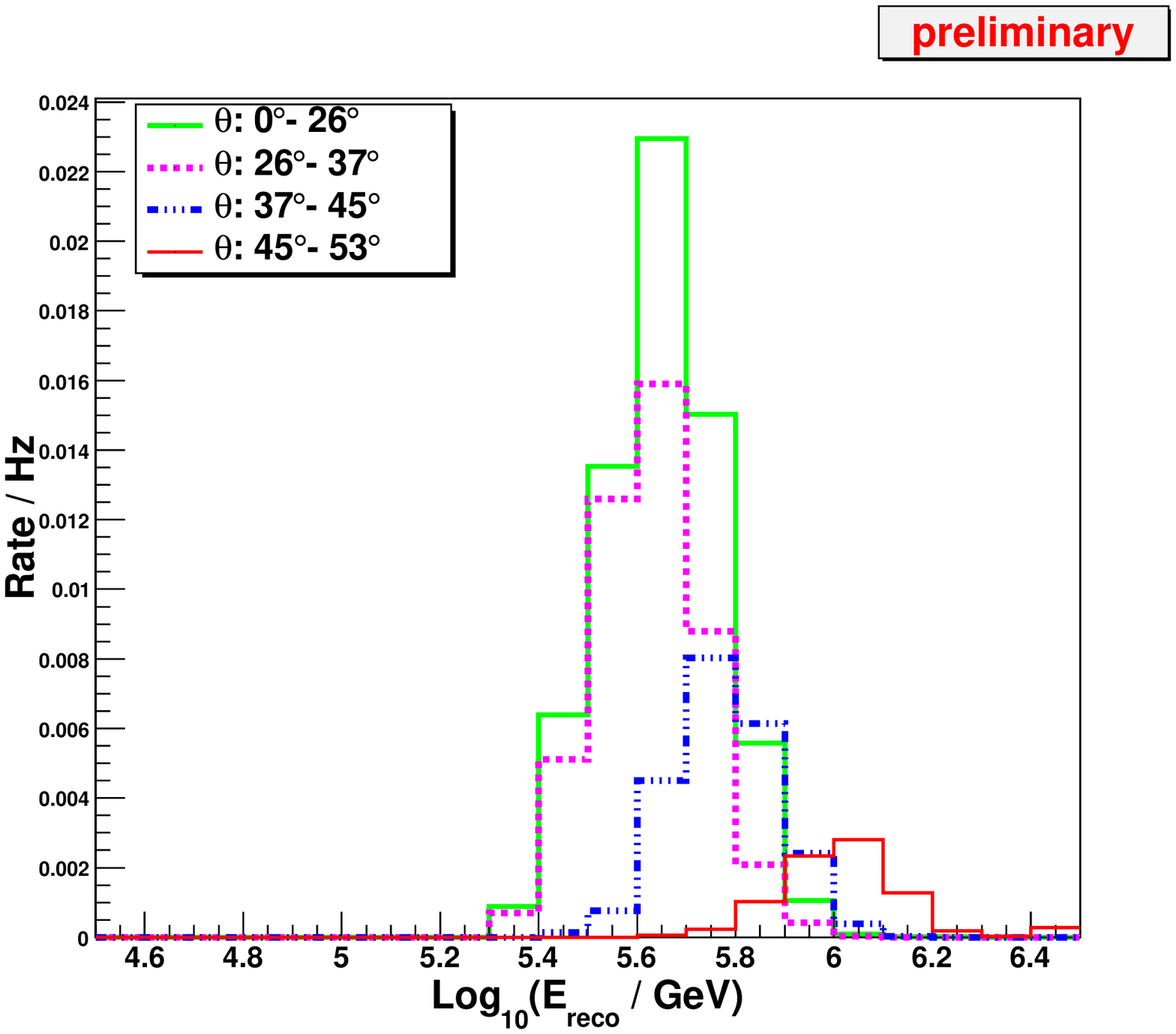} \label{sub_fig2}}
             }
   \caption{Reconstructed energy distributions for proton and iron simulations for 3 station events in four zenith bins.}
   \label{double_fig_1}
 \end{figure*}

\subsection{Experimental data and simulations}
The experimental data used in this analysis were taken during an eight hour run on September 1st, 2008. 
Two sets of air shower simulations were produced: pure proton primaries in the energy range of 21.4\,TeV--10.0\,PeV, and pure iron 
primaries in the energy range of 45.7\,TeV--10.0\,PeV. All air showers were produced in zenith angle range: $0^{\circ}\leq\theta\leq65^{\circ}$. 

Our simulation used the following flux model:
\begin{eqnarray}
    \frac{dF}{dE}&=&\Phi_{0}\left(\frac{E}{E_{0}}\right)^{-\gamma}\\
    \Phi_{0}&=&2.6\cdot10^{-4}\,\mathrm{GeV^{-1}s^{-1}sr^{-1}m^{-2}}\nonumber\\
     E_{0}&=&1\,\mathrm{TeV} \nonumber\\
    \gamma&=&2.7\nonumber
\end{eqnarray}
for both proton and iron primaries. The normalizations were chosen such that the fluxes will fit the all particle cosmic ray spectrum as shown in Fig.{~}\ref{flux}.
Simulated showers were dropped randomly in a circular area, around the center of the 40 station 
array ($X=100\,m$, $Y=250\,m$) with a radius of 600\,m.

\begin{figure*}[!t]
   \centerline{\subfloat[3 stations]{\includegraphics[width=3.1in]{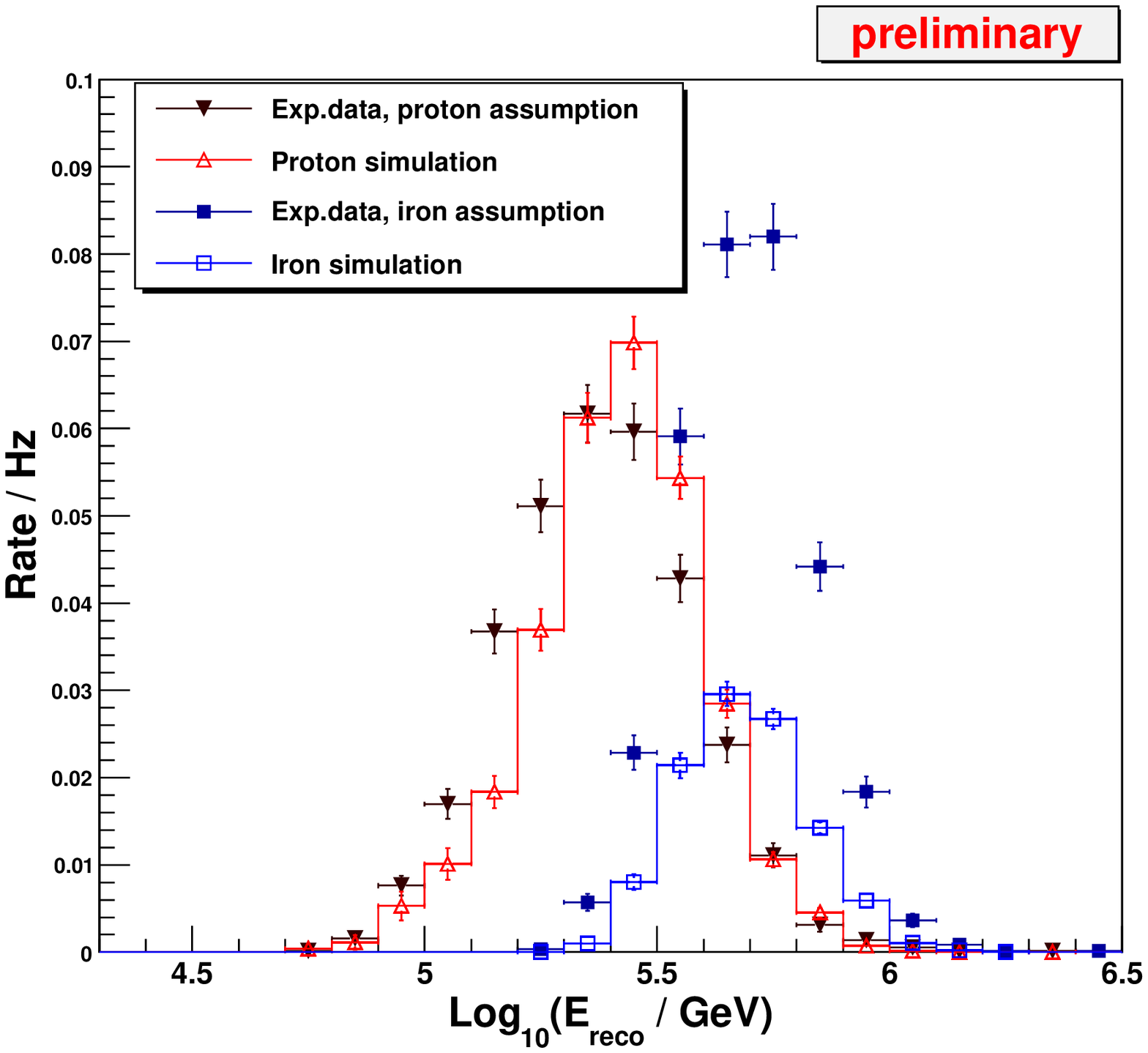} \label{sub_fig1}}
              \hfil
              \subfloat[4 stations]{\includegraphics[width=3.1in]{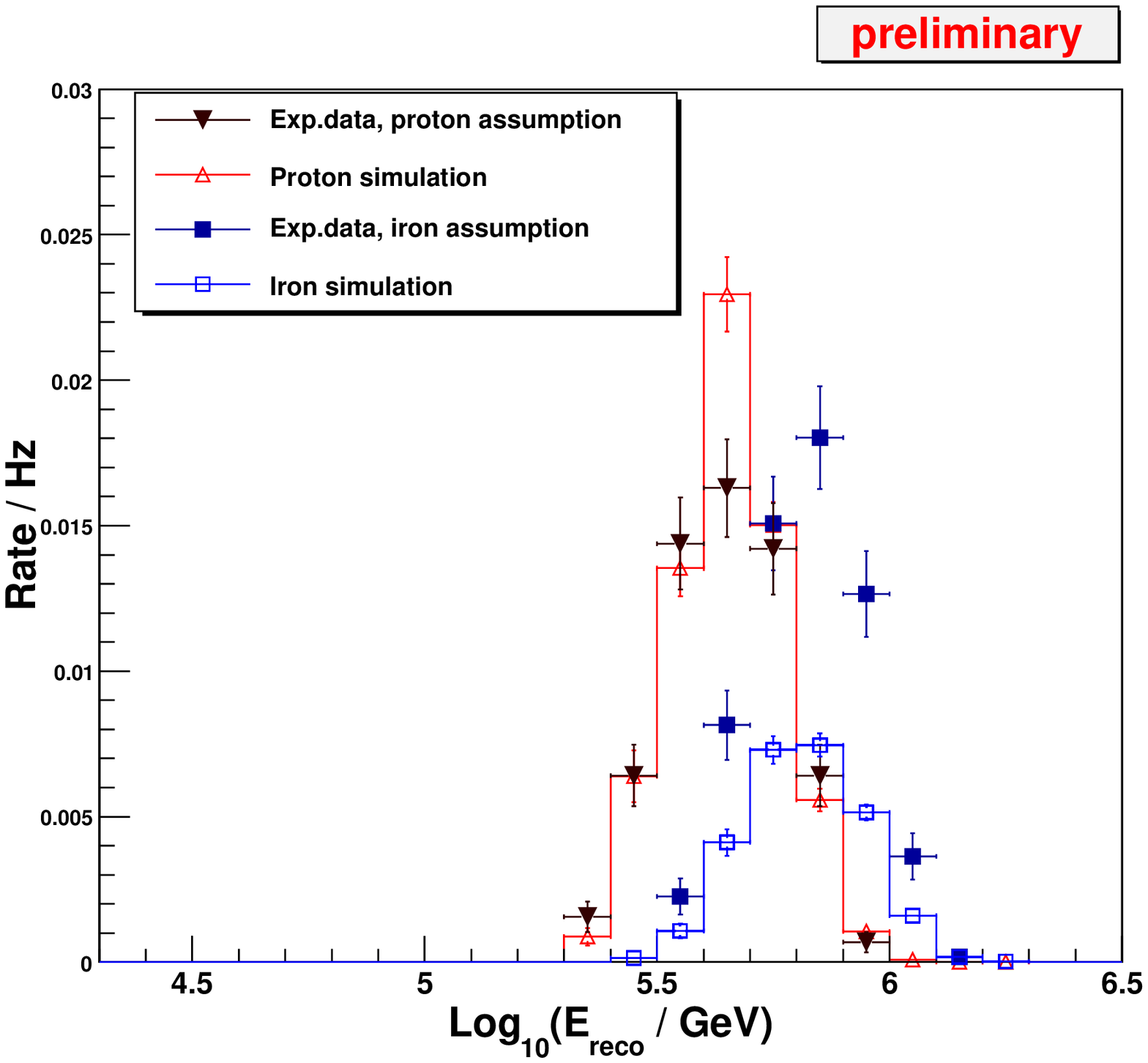} \label{sub_fig2}}
             }
   \caption{Reconstructed energy distributions for 3 and 4 station events with zenith angles $0^{\circ}$--$26^{\circ}$.
            Experimental data is reconstructed twice: assuming pure proton and pure iron primary.}
   \label{double_fig_2}
 \end{figure*}

\subsection{Shower Reconstruction}
Since the showers that trigger only three or four stations are
relatively small, we use a plane shower front approximation and the arrival times to reconstruct the
direction. The shower core location is estimated by calculating
the center of gravity of the square root of the charges in the stations. For the
energy reconstruction we use the lateral fit method \cite{Klepser} that IceTop uses to
reconstruct events with five or more stations triggered. 
This method uses shower sizes at the detector level to estimate the energy of the primary particle. 
Heavier primary nuclei produce showers that do not penetrate as deeply into the atmosphere as the proton primaries of the same energy. 
As a result, iron primary showers will have a smaller size at the detector level than proton showers of the same
energy.  We define a reconstructed energy based on simulations of primary protons and fitted to the lateral distribution
and size of proton showers. Therefore the parameter for reconstructed energy underestimates the energy when applied to showers generated by heavy primaries.
We observe a linear correlation between true and reconstructed energies in this narrow energy range and use this to correct the reconstructed energies.
We reconstruct the experimental data assuming pure proton or pure iron primaries.

\subsection{Effective Area}
We use the simulations to determine the effective area as 
a function of energy. Effective area is defined as
\begin{eqnarray}
A_{\mathrm{eff}}&=&\frac{\rm{Rate}[E_{\mathrm{min}},E_{\mathrm{max}}]}{\Delta\Omega \cdot F_{\mathrm{sum}}}\\
F_{\mathrm{sum}}&=&\int\limits_{E_{\mathrm{min}}}^{E_{\mathrm{max}}}\Phi_{0}\left(\frac{E}{E_{0}}\right)^{-\gamma}dE
\end{eqnarray}
where Rate[E$_{\mathrm{min}}$,E$_{\mathrm{max}}$] is total rate for a given energy bin, $\Delta$$\Omega$ is the solid angle of the bin and 
$\it{F}$$_{\mathrm{sum}}$ is the total flux in the given energy bin.
Figure \ref{effarea} shows the calculated effective areas, using the true values of energy and direction, for different trigger combinations, 
in the most vertical bin ($\theta = 0^{\circ}$--$26^{\circ}$).

\begin{figure}[!t]
  \centering
  \includegraphics[width=3.1in]{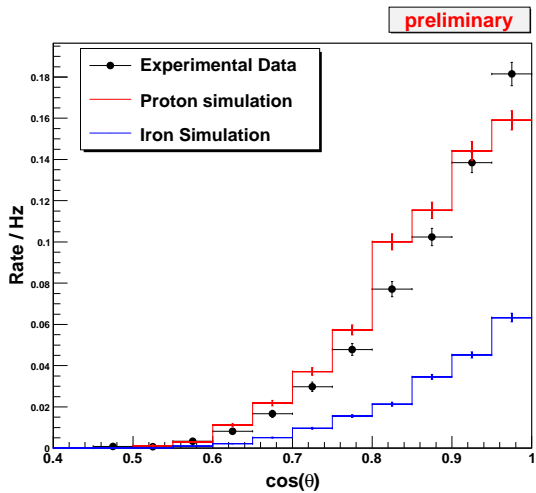}
  \caption{Reconstructed zenith distributions for 3 station events.}
  \label{cos}
 \end{figure}

\section{Results}
We summarize the results of simulations and comparison to data in Figures \ref{double_fig_1}--\ref{cos}.

In Fig.\ref{double_fig_1}, we see that the energy distribution
of the event rates depends on the zenith angle and the primary type. As
expected, the peak of the energy distribution moves to higher energies for
larger zenith angles and heavier primaries; these features of the
distributions will be very helpful in unfolding the cosmic ray spectrum and
composition.

Figure \ref{double_fig_2} shows the energy distributions in the most vertical zenith bin ($\theta = 0^{\circ}$--$26^{\circ}$). 
Experimental data is reconstructed twice, first with a pure proton assumption, then with a pure iron assumption. For three stations triggered (Fig.{~}\ref{double_fig_2}a), 
the energy distribution for pure proton simulation with the flux model as defined in (1) has a better agreement to the experimental data than iron simulation.
For four stations triggered (Fig.{~}\ref{double_fig_2}b), we have a similar picture but the peaks of the distributions are shifted to the right since on average we need a 
higher energy primary to trigger four stations. By including three station events we can lower the threshold down to 130\,TeV.

Figure \ref{cos} shows the zenith distributions of the events. Distribution for pure iron simulation is lower than for proton simulation since fewer iron
primaries reach the detector level at lower energies. The deficiency of simulated events in the most vertical bin may be due to the fact 
that we used a constant $\gamma$ of 2.7 for all energies and at these energies $\gamma$ is most probably changing continuously. In the most vertical bin 
showers must have a lower energy than showers at greater zenith angle. Starting from a lower $\gamma$ and  gradually increasing it for higher energies 
will increase events in vertical bin and decrease them at higher energies, thus improving the zenith angle distribution. It is possible to further improve the fit 
of the proton simulation to the experimental data by adjusting the parameters $\gamma$ and $\Phi_{0}$ of the model.

\section{Conclusion }
We have demonstrated the possibility of extending the IceTop analysis down to energies of 130\,TeV, low enough to overlap the direct measurements of cosmic rays.
Compared to IceTop effective area for five and more station hits, our results show a significant increase in effective area for energies between
100--300\,TeV (Fig. \ref{effarea}). We plan to include three and four station events in the analysis of coincident events to determine primary composition,
along the lines described in \cite{feusel}. Overall results of this analysis encourage us to continue and improve our analysis of small showers.

\end{document}